\documentclass[aps, prd,preprint,nofootinbib,graphicx,epsfig]{revtex4-1}
\usepackage{graphicx}
\usepackage{hyperref}
\begin{document}

\title{Equation of State of a Dense and Magnetized Fermion System}
\author{Efrain J. Ferrer}
\author{Vivian de la Incera}
\author{Jason P. Keith}
\author{Israel Portillo}
\author{Paul L. Springsteen}
\affiliation{Department of Physics, University of Texas at El Paso,
El Paso, TX 79968, USA}

\begin{abstract}
The equation of state of a system of fermions in a uniform magnetic field is obtained in terms of the thermodynamic quantities of the theory by using functional methods. It is shown that the breaking of the $O(3)$ rotational symmetry by the magnetic field results in a pressure anisotropy, which leads to the distinction between longitudinal- and transverse-to-the-field pressures. A criterion to find the threshold field at which the asymmetric regime becomes significant is discussed. This threshold magnetic field is shown to be the same as the one required for the pure field  contribution to the energy and pressures to be of the same order as the matter contribution. A graphical representation of the field-dependent anisotropic equation of state of the fermion system is given. Estimates of the upper limit for the inner magnetic field in self-bound stars, as well as in gravitationally bound stars with inhomogeneous distributions of mass and magnetic fields, are also found.

\pacs{21.65.Mn, 21.65.Qr, 26.60.Kp, 97.60.Jd}
\end{abstract}
\maketitle

\section{Introduction}
The largest magnetic fields observed in nature are associated to some of the most extreme astrophysical objects, the compact stars. For pulsars the typical magnitudes of surface magnetic fields have range $\sim 10^{12}-10^{13}$ G \cite{Pulsars}. The measured periods and spin down of soft-gamma repeaters (SGR) and anomalous X-ray pulsars (AXP), as well as the observed X-ray luminosities of AXP, indicate that a certain class of neutron stars called magnetars can have even larger magnetic fields, reaching surface values as large as $10^{14}-10^{15}$ G \cite{Magnetars}. Furthermore, if the suggestion \cite{Gamma} that these stars can be the source of gamma-ray bursts is confirmed, their magnetic fields should be even larger in order to drive a substantial Poynting flux-dominated relativistic outflow. Up to now, about ten highly magnetized neutron stars have been identified in our galaxy, but based on the population statistics of SGR it is expected that magnetars constitute about 10 $\%$ of the neutron star population \cite{Magnetar-Population}.

The existence of the strongest magnetic fields in compact stars poses the question of their origin. The simple hypothesis that a relative small magnetic field of a progenitor star can be amplified during the star's gravitational collapse due to magnetic flux conservation \cite{B-Collapse}, cannot even substantiate the high values of the surface fields in magnetars \cite{Flux-conservation}. Another generation mechanism is the so called  magnetohydrodynamic dynamo mechanism (MDM). The MDM is based on the amplification of a seed magnetic field due to the rapidly rotating plasma of a protoneutron star. It is generally accepted as the standard explanation for the origin of the magnetar's large magnetic fields. For the MDM to explain the large surface field strengths observed in magnetars, the rotational period of the protoneutron stars that originate them should be $<3 ms$. Nevertheless, this mechanism cannot substantiate all the features of the supernova remnants surrounding these objects \cite{magnetar-criticism,Xu06}. Part of the rotational energy is supposed to power the supernova through rapid magnetic braking, from where it is inferred that the supernova remnants associated with magnetars should be an order of magnitude more energetic than typical supernova remnants. However, recent calculations \cite{magnetar-criticism} indicate that their energies are similar. In addition, one would expect that when a magnetar spins down, the rotational energy output should go into a magnetized particle wind of ultrarelativistic electrons and positrons that radiate via synchrotron emission across the electromagnetic spectrum. Yet, so far nobody has detected the expected luminous
pulsar wind nebulae around magnetars \cite{Safi-Harb}. On the other
hand, some magnetars emit repeated flares or bursts of energy in the
range of $10^{42} - 10^{46} erg$ \cite{burst}. Since the emitted
energy significantly exceeds the rotational energy loss in the same period, it is natural to expect that the energy
unbalance could be supplied by the stellar magnetic field, which is
the only known additional energy source. Nonetheless, from the spin
history of these objects, there is no clear evidence of any surface
magnetic field damping \cite{spin}.

From the previous considerations it is clear that alternative mechanisms to the standard
magnetar model \cite{Magnetars} should be explored. A reasonable approach is to investigate possible microscopic mechanisms, based on the quantum phase of the core, through which a seed inner star's magnetic field can be generated and/or boosted. Some propositions in this direction already exist in the literature \cite{Phases, Flux-conservation}. Even though, to find a connection between the microscopic phase of the star's core and the astrophysical observations, other important properties of the star's core matter need first to be better understood. Along these lines, a very important problem to elucidate is the influence of a magnetic field on the star's equation of state (EoS).

Over the years, many works have been dedicated to the effects of magnetic fields in neutron (including hybrid) stars \cite{MNS} and in quark (strange) stars \cite{MQS}. However, in general, when finding the field-dependent contributions to the energy density and pressures, they did not follow a unique and consistent scheme, thereby different papers have different stands on what should be the correct field contributions to the pressure and energy. Moreover, it is known that in the presence of a magnetic field the pressure splits in two terms: transverse and parallel to the field direction, due to the breaking of the spational rotational symmetry. Nevertheless, some authors ignored the pressure anisotropy even at very strong magnetic fields, where it becomes significant. Besides, one can identify, depending on their origin, two different field contributions to the energy and pressures. One coming from the magnetized matter and the other from the Maxwell term. Despite all this, some of the previous studies do not take into account the pure field effect (coming from the Maxwell term) in the energy density and pressures, even though it is always present and in some limits it can be the dominant one.

The main purpose of this paper is to develop a systematic and self-consistent approach to treat the EoS of a magnetized system. We analyze under what conditions the pure magnetic contribution to the energy and pressures is much smaller than the matter contribution, as well as when it is self-consistent to neglect the differences between the transverse and parallel pressures (isotropic limit). We carry out our study in a theory of free fermions only interacting with an applied uniform and constant magnetic field, but the method we developed to analyze the effect of the different contributions to the EoS can easily and straightforwardly incorporate interactions.

Another outcome of our investigation is an improved estimate for the upper limit of compact stars' inner magnetic fields. Previous estimates were done assuming a gravitationally bound star with spherical and homogeneous mass distribution and a uniform magnetic field. For gravitationally bound stars we demonstrate that when the homogeneous mass density and uniform field distribution conditions are relaxed, the inner field in the high-dense core can attain values two orders of magnitude larger than previously found. We also estimate the inner magnetic fields in self-bound stars, which results of the same order than that of inhomogeneous gravitational bound stars. Using our magnetized fermion model, we calculate the threshold field that separates the isotropic and anisotropic regimes. This field turns out to be smaller than the estimated upper values of the stars' inner fields, indicating that the anisotropic effects can be relevant for the physics of the cores.

The outline of the paper is as follows. In Sec. II we estimate the upper limits for the magnetic field in self-bound and gravitationally bound compact stars. The derivation of the Maxwell and Dirac field contributions to the covariant energy-momentum tensor is reviewed in Sec. III. In Sec. IV, the quantum-statistical average of the energy-momentum tensor components are calculated using a functional method. From these results the system energy density, and the parallel and transverse pressures are obtained in terms of the thermodynamical quantities. A covariant structure for the energy-momentum tensor, in agreement with the symmetries of the magnetized many-particle system, is given in terms of the thermodynamic quantities. In Sec. V, the EoS of the magnetized fermion system is found at zero temperature and finite density. Numerical results for the energy density and pressures as functions of the magnetic field are presented and the significance of the matter and field contributions for the different ranges of densities and magnetic field strengths are discussed. Also, the threshold field for the transition between the isotropic and anisotropic pressure regimes is obtained. Our concluding remarks are stated in Sec. VI. In the Appendix the system thermodynamic potential is derived using Ritus's eigenfunction method.

\section{Compact Star's Field Estimates}\label{compat-stars-fields}
Since the interior magnetic fields of neutron stars are not directly accessible to observation, their possible values can only be estimated with heuristic methods. A direct application of the virial theorem leads to inner field estimates of order $\sim 10^{18}$ $G$ \cite{virial} for compact stars with masses $M\sim 1.4 M_\odot$ and radius $R\sim 10^{-4} R_\odot$. This derivation was done for gravitationally bound objects with uniform fields and mass density. In this Section, we show that if these conditions are relaxed the star's inner magnetic field may reach even higher values. Below we consider two possibilities: 1) self-bound objects with uniform magnetic fields, and 2) gravitationally bound objects with a physically more realistic case of inhomogeneous field and mass distributions.

\subsection{Self-bound compact stars}\label{Self-bounded}
Self-bound stars are stars made of stable $u$, $d$ and $s$ quark matter. Let us explain how this can be possible at least from a theoretical point of view. If the star's density is high enough for deconfinement, a quark matter phase may occur. Under these conditions, the up and down quarks can convert into s-quarks via weak interactions. In fact, the quark system will \textit{prefer} to do so in order to lower the Fermi energy by increasing the degeneracy. The charm, top and bottom quarks are not relevant in this analysis because their masses are much larger than the strange's and the typical densities of the stars will not be enough to produce these other flavors. The thereby formed three-flavor quark matter, composed of a mix of $u$, $d$ and $s$ quarks, is known as strange quark matter. It has been conjectured \cite{Bodmer}, \cite{Witten} that at zero pressure the strange quark matter will have a lower energy per baryon than ordinary matter, which has $\varepsilon_{Matter}=939$ $MeV/baryon$. This possibility would make the strange matter the most stable substance in nature. According to this, ordinary nuclei would lower their energy by converting to strange matter, but it has been shown that the conversion rate is negligible under almost all conditions, except perhaps in neutron stars \cite{Farhi}.  Therefore, the
possible existence of strange stars cannot be ruled out \cite{Witten, Farhi}.

Assuming there are strange stars out there, a reasonable question to ask is: how big a magnetic field can be sustained by them? Energy-conservation arguments can help to estimate the maximum field strength; one expects that the magnetic energy density should not exceed the energy density of the self-bound quark matter. Based on this, the maximum field allowed is estimated as
\begin{equation}\label{Bind-Energy}
B_{max}\simeq \frac{\varepsilon_{Matter}^2}{e\hbar c}\leq \frac{(939 MeV)^2}{e\hbar c}\sim1.5\times10^{20} G,
\end{equation}
We call attention that it results two orders of magnitude larger than the estimates done for gravitationally bound stars assuming uniform field and mass density.

\subsection{Gravitationally bound stars with inhomogeneous mass and field distributions}\label{Gravitational-bounded}
Let us consider now the case of gravitationally bound stars. As known, neutron stars are the remnant of type-II supernova explosions. In a neutron star, pressure rises from zero (at the surface) to an unknown large value in the center.   Also the density changes from surface values much smaller than the saturation density, $\rho_s\approx 4\times10^{11}$ $g/cm^3$ -the density at which nuclei begin to touch- to inner values several times the normal nuclear density $\rho_N=2.8\times 10^{14}$ $g/cm^3$. At such high densities, deconfinement can occur and the star's core can have quark matter. Thus, the cores of very massive neutron stars are the best natural candidates for the realization of the transition from hadronic matter to a deconfined quark phase. This possibility was pointed out by several authors long time ago \cite{Quark-stars}. Stars with radius-dependent density leading to confined nuclear matter in the outer region and deconfined quark matter in the core are called hybrid stars. For excellent reviews on this topic see \cite{NS-Reports}.

What can we say about the star's magnetic field in this case? Well, we know that the stellar medium has a very high electrical conductivity, hence the magnetic flux is
conserved. Since the flux is conserved, during the formation of the neutron star, that is, during the protoneutron star period, the magnetic field strength should increase with the increase of the matter density. In addition, some of the proposed phases \cite{Phases}-\cite{PMCFL} that could be realized in the core of the compact object that result at the end of the supernova explosion might also contribute to increase
the field strength even more. For instance, if the resulting core is dense enough as to be in the color superconducting color-flavor-locked (CFL) phase and a magnetic field of strength comparable to the Meissner mass of the "rotated" charged gluons were present, this field would trigger an instability \cite{PMCFL} which in turn would lead to the generation of a gluon vortex condensate. The gluon vortex state so produced behaves as a paramagnet, thus increasing the total magnetic
field in the core. This vortex phase is known in the literature as the paramagnetic CFL (PMCFL) phase \cite{PMCFL}. As these arguments indicate, the magnetic field in the core could be higher than in the surface.

Let us briefly outline the field estimate obtained in Ref. \cite{virial} for the case of homogenous field and mass distributions.  From the equipartition theorem under these conditions
\begin{equation}\label{Uniform-Field}
\left(\frac{4}{3}\pi R^3\right)\frac{H^2}{8\pi}=\frac{3}{5}G\frac{M^2}{R},
\end{equation}
they found that the maximum field was given by
\begin{equation}\label{Max-Uniform-Field}
H_{max}=H_\odot \left(\frac{M}{M_\odot}\right) \left(\frac{R}{R_\odot}\right)^{-2},
\end{equation}
Using $H_\odot =2\times 10^8$ $G$, and taking into account that a typical neutron star has $M\simeq1.4 M_\odot$ and $R\simeq (0.14\times10^{-4}) R_\odot$, one easily
estimates the maximum strength as $H_{max}\sim10^{18}$ $G$.

But as we said above, we are interested in the more realistic situation where the constraints of uniform mass and field distributions are relaxed. With this aim in mind, let us assume that both the mass density and the magnetic field increase from the surface $(r=R)$ to the star center $(r=0)$ and let us consider the following parametrizations for the density and magnetic field respectively
\begin{equation}\label{Inhomogeneous-Mass-1}
\rho=\rho_0\left[1-\left(\frac{r}{R}\right)^a \right],\qquad a>0,
\end{equation} \begin{equation}\label{Inhomogeneous-Field}
H(r)=H_S+\left(H_0-H_S \right)\frac{R^b-r^b}{R^b},\qquad b>0,
\end{equation}
$\rho_0$ is the density at the core; $H_s$ and $H_0$ are the magnetic fields at the surface and inner core respectively; and $a$ and $b$ are parameters to be determined.

Using the mass density (\ref{Inhomogeneous-Mass-1}), a star with a spherical configuration of radius $R$ will have mass
\begin{equation}\label{Inhomogeneous-Mass}
M=\rho_0V\left(\frac{a}{a+3}\right),
\end{equation}

The coefficient $a$ must be positive, but apart from that it is totally arbitrary. We can use typical values of neutron stars' mass $M=1.4M_\odot$ and
radius $R=10 km$, as well as realistic core density estimates to find a region of physically acceptable values for $a$. The results are shown in
Fig.1-(a). The plot gives the mass density coefficient $a$ as a function of the parameter $n$ that characterizes how much larger than the nuclear density is
the star's core density $\rho_0=n\rho_N$, $n=1,2,...$. Notice that $n$ must be larger than 2 to obtain an acceptable (positive) value of $a$. For $n=3$, a realistic core density case, the parameter $a$ is positive and lies between 10 and 20.

\begin{figure}
\begin{center}
\includegraphics{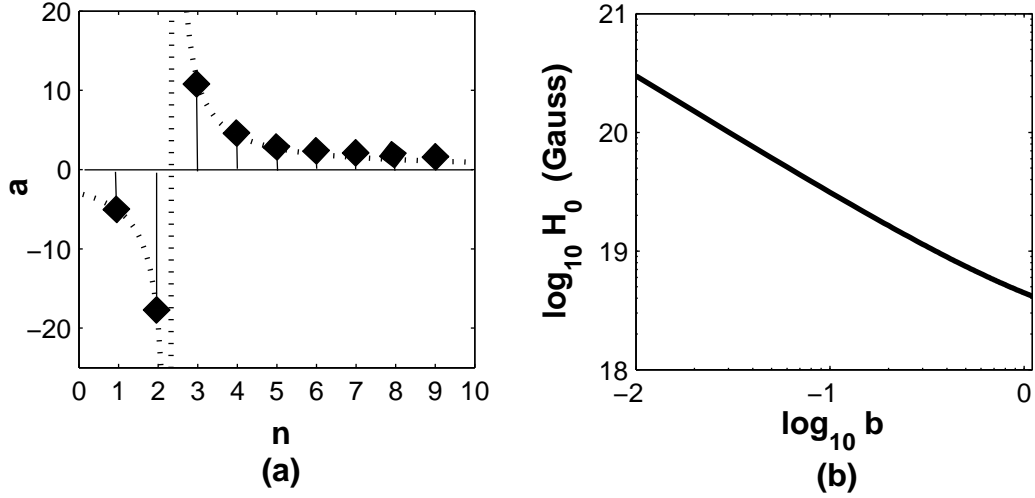}
\caption{(a) The mass density coefficient $a$ vs the density parameter $n$. (b) The core magnetic field $H_0$ vs the field coefficient $b$.} \label{fig1}
\end{center}
\end{figure}

As is $a$, the parameter $b$ is also arbitrary. To obtain a reasonable set of values for $b$ we need to use the equipartition theorem in our inhomogeneous star model. The gravitational energy of a spherical distribution of mass with density (\ref{Inhomogeneous-Mass-1}) and radius R is
\begin{equation}\label{Grav-Energy}
W_g=2\pi\int_0^R \rho(r)\phi(r)r^2dr=-4\pi^2G\rho_0^2R^5F(a)
\end{equation}
with $F(a)$ given by
\begin{equation}\label{F-a}
F(a)=\frac{8a^4+60a^2+87a^2}{15(a+2)(a+3)(a+5)(2a+5)},
\end{equation}
The gravitational potential $\phi(r), r\leq R$ in (\ref{Grav-Energy}) is
\begin{eqnarray}\label{Potential}
\phi(r)\equiv-G\int\frac{\rho(r')dV'}{|\overrightarrow{r}-\overrightarrow{r}'|}=\qquad\qquad\qquad\qquad\qquad\qquad\qquad \quad\quad \nonumber
\\
=-2\pi G\rho_0R^2 \left[\frac{a}{a+2}-\frac{1}{3}\left(\frac{r}{R}\right)^2+\frac{2}{(a+2)(a+3)}\left(\frac{r}{R}\right)^{a+2}\right ]
\end{eqnarray}

On the other hand, the magnetic energy corresponding to the field configuration (\ref{Inhomogeneous-Field}) is
 \begin{eqnarray}\label{Potential}
W_m\equiv 4\pi \int_0^R\frac{H^2(r)}{8\pi}r^2dr=\frac{1}{2}\int_0^R \left[H_0-(H_0-H_s)\left(\frac{r}{R}\right)^b\right]^2r^2dr \quad\qquad   \nonumber
\\
=\left[\frac{b^2}{3(b+3)(2b+3)}H_0^2+\frac{b^2}{b(b+3)(2b+3)}H_0H_s+\frac{1}{2(2b+3)}H_s^2\right ]R^3
\end{eqnarray}

Taking into account that magnetars' surface fields are $\sim
H_s=10^{14} G$, and considering the value of $a$ that corresponds to
$\varrho_{0}=3\varrho_{N}$ \footnote{Densities of this order are
large enough for deconfinement.}, we can use the equipartition of
the magnetic (\ref{Potential}) and gravitational (\ref{Grav-Energy})
energies, $W_g=W_m$, to graphically find the inner field $H_0$ as a
function of the parameter $b$. The resulting curve is plotted in
Fig. 1-(b). The smaller $b$ is, the slower the
field decays from the inner region to the surface, and thus the larger the
inner field required to produce $H_s=10^{14} G$ in the surface. For a value of $b$ between $0.1$ and $0.01$ the core field can be larger than ($10^{19}$ $G$).

The above estimate for the field is model-dependent. Even though we make no claim that this ad-hoc model correctly describes the real way the field varies with the radius in a hybrid star, our derivation serves to illustrate how the more realistic assumption of inhomogeneity can be consistent with stronger field estimates in the core than those previously found in the literature using homogeneous distributions. Since at present there is no reliable way to know the exact dependence of the density and field with the radius in a real neutron star, we will have to wait for more observations to validate or not this possibility.

We will see now, based on energy conservation arguments within a microscopic analysis, that there is a natural scale for the core magnetic field $\sim 10^{20}$ $G$. Given the energy density
per baryon at the core of the gravitationally bound system
\begin{equation}  \label{Energy-baryon}
\frac{\epsilon}{n_A}=-\frac{p_\|}{n_A}+\mu \frac{N} {n_A},
\end{equation}
where the baryon number $n_A=\frac{1}{3}(n_u+n_d+n_s)=N/3$, let us assume that there will be some field value from where the parallel pressure becomes negligible ($p_\|\simeq
0$) and then let us estimate the order of this maximum field just by reasoning that the magnetic energy density should be at most as large as the energy density of the baryon system.  Then, neglecting the first term in the RHS of Eq. (\ref{Energy-baryon}), one has
\begin{equation}  \label{Energy-baryon-2}
\frac{\epsilon}{n_A}=3\mu
\end{equation}
and consequently,
\begin{equation}  \label{Estimated-H}
\widetilde{H}\simeq 9\mu^2/e\hbar c
\end{equation}
For the phenomenologically acceptable value of $\mu\simeq 400$ $MeV$, we get from (\ref{Estimated-H}) that the
magnetic field $\sim 10^{20}G$. As will be shown later in the paper, the parallel pressure indeed decreases with the field and for a system of free fermions becomes negligibly small at a field strength of order $10^{19}G$.

Thus, we conclude that field strength of $~10^{20}$ $G$ makes a natural maximum scale for both self-bound and at the core of high-density inhomogeneous gravitational-bound compact stars.

\section{Energy-Momentum Tensor at H$\neq$ 0}\label{energy-momentu-tensor}

For the sake of completeness and understanding, let us outline the steps that lead to a symmetric and gauge invariant energy-momentum tensor of matter and fields. The energy-momentum tensor $T_{\mu\nu}$ is associated with the Noether's currents of an infinitesimal space-time-dependent translation and in terms of the Lagrangian density $\cal L(\varphi,\partial_\mu\varphi)$ of the theory it is given by
\begin{equation}\label{Ene-Mom}
T^{\mu\nu}\equiv\frac{\partial{\cal L}}{\partial(\partial_\mu\varphi)}\partial^\nu\varphi-\cal L\delta^{\mu\nu}
\end{equation}
This $T^{\mu\nu}$ tensor is the so-called canonical energy-momentum tensor. In general it is neither symmetric non gauge invariant. Nevertheless, these two shortcomings can be fixed by taking advantage of the arbitrariness of $T^{\mu\nu}$, which is defined up to a total derivative
\begin{equation}\label{Ene-Mom-Amboguity}
T^{\mu\nu}\rightarrow T^{\mu\nu}+\partial_\rho M^{\mu\nu\rho}
\end{equation}
with $M^{\mu\nu\rho}=-M^{\rho\nu\mu}$. The redefinition (\ref{Ene-Mom-Amboguity}) preserves the conservation $(\partial_\mu T^{\mu\nu}=0)$ and the value of the global energy-momentum four-vector
\begin{equation}\label{Ene-Mom-Vector}
P^{\mu}=\int d^3x T^{\mu 0}=\int d^3x (T^{\mu 0}+\partial_\rho M^{\mu 0 \rho}),
\end{equation}
as long as $M^{i0\mu}$ vanishes sufficiently rapidly at infinity. Then, the energy and momentum, which are measurable quantities, as well as their conservations, are maintained despite the addition of a new tensor to the canonical one. With a suitable choice of the tensor $M^{\mu\nu\rho}$ the energy-momentum tensor can be converted into a symmetric and gauge invariant one.

On the other hand, there is a general derivation which guarantees from the beginning the symmetry and gauge invariance of $T^{\mu\nu}$ (denoted by $\tau^{\mu \nu}$ from now on). The idea derives from the fact that if the matter fields are coupled to gravity, the energy-momentum tensor plays the role of the source of the gravitational field. In this case, the introduction of gravity will generalize the space-time transformations to the frame of the general covariance of general relativity, rather than to the particular Lorentz transformations. Thus, the program to follow is to consider the starting theory in a curved space-time geometry where the Minkowski metric $\eta_{\mu\nu}$ is replaced with the general metric $g_{\mu\nu}$, the volume element $d^4x$ by $d^4x\sqrt{-g}$ (with $g$ being the determinant of the metric), and then to obtain the energy-momentum tensor through the invariance of the action with respect to the variation of the metric \cite{Landau-Lifshitz}
\begin{equation}\label{Ene-Mom-Tensor-Derivat}
-\frac{1}{2}\int d^4x \sqrt{-g}\delta g_{\mu\nu}\tau^{\mu \nu}=0
\end{equation}
Once a manifestly symmetric and gauge-invariant $\tau^{\mu\nu}$ tensor is obtained by this procedure we can switch off the gravitational field by returning to the Minkowski metric ($g_{\mu \nu} \rightarrow \eta_{\mu \nu}$).

To find the energy-momentum tensor of a fermion system in the presence of a magnetic field, we start from the Lagrangian density
\begin{equation}\label{Lagrangian}
\cal {L}(\psi, \textsl{F}_{\mu\nu})=\cal {L}_{A_\mu}(\textsl{F}_{\mu\nu})+{\cal L}_\psi(\psi, \textsl{F}_{\mu\nu})
\end{equation}
where $\cal {L}_{A_\mu}(\textsl{F}_{\mu\nu})$ denotes the Maxwell Lagrangian density and ${\cal L}_\psi(\psi, \textsl{F}_{\mu\nu})$ that for the Dirac field in the presence of the external magnetic field. They are respectively given by
\begin{equation}\label{Maxwell-Lagrangian}
\cal {L}_{A_\mu}(\textsl{F}_{\mu\nu})=-\frac{\textsl{1}}{\textsl{4}}\textsl{F}_{\mu\nu}\textsl{F}^{\mu\nu}
\end{equation}
and
\begin{equation}\label{Dirac-Lagrangian}
{\cal L}_\psi(\psi, \textsl{F}_{\mu\nu})=\frac{1}{2}\overline{\psi}(\overrightarrow{D}_\mu \gamma^\mu-m)\psi+\frac{1}{2}\overline{\psi}(\overleftarrow{D}_\mu \gamma^\mu -m)\psi
\end{equation}
with the right and left gauge covariant derivatives given respectively by
\begin{equation}
\overrightarrow{D}_{\mu }=i\overrightarrow{\partial} _{\mu }-eA_{\mu }  \label{Covariant-Derv-R}
\end{equation}
and
\begin{equation}
\overleftarrow{D}_{\mu }=-i\overleftarrow{\partial} _{\mu }-eA_{\mu }  \label{Covariant-Derv-L}
\end{equation}
where $A_{\mu }$ is the electromagnetic potential associated with the external applied field.

\subsection{Energy-momentum tensor of the Maxwell field}\label{T-Maxwell}

To find the manifestly symmetric and gauge invariant energy-momentum tensor for the Maxwell field we have from (\ref{Ene-Mom-Tensor-Derivat}) that
\begin{equation}\label{T-M}
 \tau^{\rho \lambda}_{A_\mu}= \frac{-2}{\sqrt{-g}}\frac{\delta}{\delta g_{\rho \lambda}}(\sqrt{-g}\widetilde{\cal {L}}_{A_\mu})
\end{equation}
where $\widetilde{\cal {L}}_{A_\mu}$ is obtained from (\ref{Maxwell-Lagrangian}) by explicitly introducing the dependence on the metric tensor
\begin{equation}\label{Maxwell-Relat}
\widetilde{\cal {L}}_{A_\mu}=-\frac{1}{4}F_{\mu \nu}F_{\rho \lambda}g^{\mu \rho}g^{\nu \lambda}
\end{equation}

Taking into account that
\begin{equation}\label{Variation-1}
\frac{\delta \sqrt{-g}}{\delta g_{\rho \lambda}}   =\frac{1}{2}\sqrt{-g} g^{\rho \lambda}
\end{equation}

\begin{equation}\label{Variation-2}
\frac{\delta g^{\mu \nu}}{\delta g_{\rho \lambda}}   =- g^{\rho \mu} g^{ \lambda \nu}
\end{equation}
one finds
\begin{eqnarray}\label{T-M-variation}
 \tau^{\mu \nu}_{A_\mu}= \frac{1}{2\sqrt{-g}}\frac{\delta}{\delta g_{\mu \nu}}\sqrt{-g}F_{\sigma \tau}F_{\rho \lambda}g^{\sigma \rho}g^{\tau \lambda}\nonumber
 \\
 =-F^{\mu \rho} F^{\nu}_\rho - \widetilde{\cal {L}}_{A_\mu} g^{\mu \nu} \qquad \qquad \quad
\end{eqnarray}

Returning to the Minkowski space and considering in particular the case for a constant and uniform magnetic field $H$ along the $x_3$-direction we have from (\ref{T-M-variation}) that
\begin{equation}\label{Tau}
\tau^{\mu \nu}_{M}=(\varepsilon -p)u^\mu u^\nu +p(\eta_{\|}^{\mu \nu} - \eta_{\perp}^{\mu \nu})
\end{equation}
where $\varepsilon =p=H^2/2$, $u_\mu$ is the medium four-velocity which in the rest system takes the value $u_\mu =(1,\overrightarrow{0})$, $\eta_{\|}^{\mu \nu}$ is the longitudinal Minkowskian metric tensor with $\mu, \nu =0,3$ and $\eta_{\perp}^{\mu \nu}$ is the transverse Minkowskian metric tensor with $\mu, \nu =1,2$. The fact that the energy-momentum tensor of the magnetized space becomes anisotropic, having different pressures in the longitudinal $p_\|$ and transverse $p_\perp$ directions ($-p_\|=p_\perp=H^2/2$), is due to the breaking of the rotational symmetry $O(3)$ produced by the external field. As a consequence, the Minkowskian metric splits in two structures, one transverse $\eta_{\perp}^{\mu \nu}=\widehat{F}^{\mu \rho}\widehat{F}_{\rho}^\nu$ (where $\widehat{F}^{\mu \rho}=F^{\mu \rho} /H$ denotes the normalized electromagnetic strength tensor) and another longitudinal $\eta_{\|}^{\mu \nu}=\eta^{\mu \nu}-\widehat{F}^{\mu \rho}\widehat{F}_{\rho}^\nu$.

\subsection{Energy-momentum tensor of the Dirac field}\label{T-Dirac}

To follow the previous approach in the case of spinor fields is more involved. The problem is that there is no representation of the $GL(4)$ group of general relativity which behaves like spinors under the Lorentz subgroup \cite{Weinberg}. Then, to put fermions in interaction with a gravitational field a new formalism is required. The first formulation of spinor fields in Riemannian space-time was done in \cite{Fock} by introducing the so called vierbein or tetrad fields \cite{Weldon}. The vierbeins $V^\mu_\alpha(x)$ connect the Minkoskian metric $\eta_{\mu \nu}$ with the metric tensor of a general coordinate system $g_{\alpha \beta}$ by the map
\begin{equation}\label{Metrics}
    g_{\alpha \beta}=V^\mu_\alpha(x)V^\nu_\beta(x)\eta_{\mu \nu}\,
\end{equation}
where
\begin{equation}\label{vierbein}
    V^\mu_\alpha(X)\equiv \left(\frac{\partial y^\mu_X}{\partial x^\alpha}\right)_{x=X}
\end{equation}
with $y^\mu_X$ being the normal coordinates of a local Minkowski space at point $X$, and $x^\alpha$ the corresponding general coordinates at that point. In this way, the geometries in general relativity can be described in terms of a vierbein field instead of the usual metric tensor field. We have adopted the convection that the indices given by Greek letters at the beginning of the alphabet (i.e. $\alpha$, $\beta$, etc.), are related to magnitudes in the general reference system, while those at the end (i.e. $\mu$, $\nu$, $\rho$, etc.) to magnitudes in the local Minkowsky reference system.

On the other hand, the introduction of the vierbein fields also make it possible to generalize the algebra of the Dirac gamma-matrices $\gamma^\mu$, given by the anticommutation relation
\begin{equation}\label{gamma-rel-MS}
    \left\{\gamma^\mu,\gamma^\nu \right\}=2\eta^{\mu \nu}I
\end{equation}
to the curved space
\begin{equation}\label{gamma-rel-CS}
    \left\{\gamma^\alpha,\gamma^\beta \right\}=2g^{\alpha \beta}I
\end{equation}
with $\gamma^\alpha=V^\alpha_\mu(x)\gamma^\mu$.

The Lagrangian density of the Dirac fields (\ref{Dirac-Lagrangian}) can be rewritten in curved space-time with the aid of the vierbeins fields as
\begin{equation}\label{Dirac-Lagrang-CS}
  \widetilde{\cal {L}}_\psi=\frac{1}{2} \left[\overline{\psi} \gamma^\mu V_\mu ^\alpha V_\alpha^\nu  \nabla_\nu \psi -\left( \nabla_\nu \overline{\psi} \right )V_\alpha ^\nu V_\mu ^\alpha \gamma^\mu \psi \right]-m\overline{\psi}\psi
\end{equation}
where $\nabla_\nu =V_\nu ^\alpha (D_\alpha+\Gamma_\alpha)$ is the covariant derivative in curved space with connection $\Gamma_\alpha (x)=\frac{1}{2}\Sigma ^{\mu \nu} V_\mu^\beta (x) (\frac{\partial}{\partial x^\alpha} V_{\beta \nu} (x) )$ and $\Sigma ^{\mu \nu}= \frac{1}{4}\left [ \gamma^\mu , \gamma^\nu \right ]$ being he generator of the Lorentz group.

Taking into account that
\begin{equation}\label{Aux-Eqs}
\sqrt{-g}=\det [V(x)], \qquad \delta g_{\alpha \beta} =-  \left ( g_{\alpha \gamma} V^\mu_\beta + g_{\beta \gamma} V^\mu_\alpha \right ) \delta V_\mu^\gamma
\end{equation}
it is found that the energy-momentum tensor obtained from (\ref{Dirac-Lagrang-CS}) is \cite{Davies}
\begin{eqnarray}\label{Ene-Mom-Dirac}
\tau^{\alpha \beta}_{\psi}(x)=\frac{V^{\mu \alpha}(x)}{\det [V(x)]}\frac{\delta \det [V(x)] \widetilde{\cal {L}}_\psi}{\delta V_{\beta}^{\mu}} \qquad \quad \qquad \quad \nonumber
\\
=\frac{i}{2}\left [\overline{\psi}\gamma^{( \alpha}\nabla ^{\beta )} \psi- (\nabla ^{( \alpha}\overline{\psi})\gamma^{\beta )}\psi \right ]+g^{\alpha \beta}\overline{\psi}m \psi
\end{eqnarray}

Returning to Minkowski space (i.e. with the replacement $\gamma^\alpha \rightarrow \gamma^\mu, \nabla_\alpha \rightarrow D_\mu$) we obtain
\begin{equation}\label{Ene-Mom-Dirac-Minkowski}
\tau^{\mu \nu}_{\psi}(x)=\frac{1}{2}\overline{\psi}\left [\gamma^\mu \overrightarrow{D}^\nu+\gamma^\nu \overleftarrow{D}^\mu\right ]\psi- \eta^{\mu \nu}\overline{\psi}\left [iD_\rho \gamma^\rho -m \right ]\psi
\end{equation}

Let's assume a uniform and constant magnetic field $H$ along the $x_3$-direction and use the Landau gauge $A_{\mu }^{\mathit{ext}} = H x_1 \delta_{\mu 2}$ in the covariant derivative $D_\mu \equiv \partial_\mu +ie A_{\mu }$. Then, we find
\begin{equation}\label{T-zero}
\tau^{00}_\psi(x)=\frac{1}{2}\overline{\psi}i\partial^0 \gamma^0\psi-\frac{1}{2}\psi i\partial^0 \gamma^0\overline{\psi}-\cal{L}_\psi,
\end{equation}

\begin{equation}\label{T-Longitudinal}
\tau^{33}_\psi(x)=\overline{\psi}i\partial^3 \gamma^3\psi+\cal{L}_\psi,
\end{equation}
and
\begin{equation}\label{T-Transverse}
\tau^{jj}_\psi(x)=i\overline{\psi}(D^j \gamma^j)\psi+{\cal{L}_{\psi}}, \qquad \textit{j}=1,2
\end{equation}

Once again, the asymmetry between the longitudinal and transverse diagonal components of $\tau^{\mu \nu}_{\psi}$, is related to the breaking of the $O(3)$ symmetry. Moreover, since the Landau gauge is not symmetric, we have that although a magnetic field along the $x_3$-direction conserves the rotation group $O(2)$ in the corresponding perpendicular plane, the asymmetry of the potential introduces an apparent asymmetry in the transverse indices (i.e. $\tau^{11}_{\psi}\neq \tau^{22}_{\psi}$). This apparent asymmetry, as we will see in the following section, cannot be present in the quantum-statistical average of $\tau^{\mu \nu}_{\psi}$, which can only depends on the field and not on the potential.

\section{\textsc{Energy and pressures of the dense magnetized system}}\label{Energy-P}

As is known \cite {Landau-Lifshitz}, in the reference frame comoving with the many-particle system, the system normal stresses (pressures) can be obtained from the diagonal spatial components of the average energy-momentum tensor $\langle \tau^{ii} \rangle$, the system energy, from its zeroth diagonal component $\langle \tau^{00} \rangle$, and the shear stresses (which are absent for the case of a uniform magnetic field) from the off-diagonal spatial components $\langle \tau^{ij} \rangle$. Then, to find the energy density, and pressures of the dense magnetized system we need to calculate the quantum-statistical averages of the corresponding components of the energy-momentum tensor of the fermion system in the presence of a magnetic field.

 These calculations were carried out long time ago in Ref. \cite{Canuto}, where it was used a QFT second-quantized approach. There, a quantum-mechanical average of the energy-momentum tensor  in the eigen-states of the Dirac equation in the presence of the uniform magnetic field was first performed to get the corresponding quantum operator in the occupation-number space. The macroscopic stress-energy tensor was then found by averaging its quantum operator, in the statistical ensemble using the many-particle density matrix. In this Section we perform similar calculations, but using a functional-method approach that makes easier to recognize the thermodynamical quantities entering in the final results. Our procedure is also different from that of Ref. \cite{Canuto} in the sense that we will not assume that the fermion fields entering in the definitions of the energy and pressures satisfy the classical equation of motions (i.e. the Dirac equations for $\psi$ and $\overline{\psi}$), but the functional integrals integrate in all field configurations. Then, we keep the terms depending on the Lagrangian density $\cal{L}_\psi$ in Eqs. (\ref{T-Longitudinal})-(\ref{T-Transverse}), while in Ref. \cite{Canuto} the condition $\cal{L}_\psi$$=0$ was considered.

 The quantum-statistical average of the energy-momentum tensor is given by
\begin{equation}\label{T-average}
\langle \widetilde{\tau}^{\rho \lambda} \rangle=\frac{Tr\left[\widetilde{\tau}^{\rho\lambda}e^{-\beta(H-\mu N)}\right]}{Z}
\end{equation}
where
\begin{equation}\label{T-average-2}
\widetilde{\tau}^{\rho \lambda}=\int_0^\beta d\tau\int d^3x  [\tau_{A_\mu}^{\rho\lambda}+\tau_\psi^{\rho\lambda}]
\end{equation}
and $Z$ is the partition function of the grand canonical ensemble given by
\begin{equation}\label{Partition-function}
Z=Tre^{-\beta(H-\mu N)}
\end{equation}
with $H$ denoting the system Hamilonian, $N$ the particle number, $\beta$ the inverse absolute temperature, and $\mu$ the chemical potential. The partition function (\ref{Partition-function}) can be written as a functional integral \cite{Kapusta}
\begin{equation}\label{PF-Functional-Int}
Z=\int [d\phi]e^{\int_0^\beta d \tau \int d^3 x {\cal{L}} (\tau , x)}
\end{equation}
where ${\cal{L}} (\tau , x)$ is the many-particle system Lagrangian density (i.e. with $\mu \neq 0$).  The chemical potential enters in ${\cal{L}} (\tau , x)$ as a shift in the zero component of the electromagnetic potential $A_0\rightarrow A_0-\frac{\mu}{e}$ \cite{chemical potential}.

\subsection{Energy Density}
To calculate the system energy, $\langle \widetilde{\tau}^{00}_{M}+\widetilde{\tau}^{00}_\psi \rangle$, we should notice that the external applied magnetic field behaves as a classical field in this formalism. Thus, $\langle \widetilde{\tau}^{00}_{M} \rangle =\beta V\frac{H^2}{2}$, being the pure Maxwell contribution, with $V$ denoting the system volume.

To get the fermion contribution to the energy we will make use of the functional integral. We start by calculating the quantum-statistical average
\begin{equation}\label{T00-average}
\langle \widetilde{\tau}^{00}\rangle=\frac{\int[d\psi][d\overline{\psi}]\widetilde{\tau}^{00}e^{  \int_0^\beta d\tau \int d^3x {\cal{L}_\psi}(\tau,x)}}{Z}
\end{equation}
where
\begin{equation}\label{T00-average-1}
\widetilde{\tau}^{00}=\int_0^\beta d\tau \int d^3x \tau^{00}(\tau,x)
\end{equation}
and
\begin{equation}\label{Lagrangian}
{\cal{L}}_\psi=\overline{\psi}\left[i\gamma^0(\partial^0-i\mu)-i\gamma^1\partial^1-i\gamma^2(\partial^2+ieHx^1)-i\gamma^3\partial^3-m\right]\psi
\end{equation}
is the many-particle Lagrangian density. Notice that in (\ref{T00-average}) we are not integrating in the photon field $A_\mu$, since we are taking the Maxwell field as an external classical field. Then, to find the fermion contribution to the energy density we have to calculate $\frac{1}{\beta V} \langle \widetilde{\tau}_\psi^{00}\rangle$,
with $\widetilde{\tau}^{00}_\psi$ given from (\ref{T-zero}) as
\begin{eqnarray}\label{T00-bar}
\widetilde{\tau}_\psi^{00}=\int_0^\beta d\tau \int d^3x \left(i\overline{\psi}\gamma^0\partial^0\psi\right)+\qquad\qquad\qquad\qquad\qquad\qquad\qquad\quad\nonumber
 \\
 -\int_0^\beta d\tau \int d^3x \overline{\psi}\left[i\gamma^0\partial^0-i\gamma^1\partial^1-i\gamma^2(\partial^2+ieHx^1)-i\gamma^3\partial^3-m\right]\psi
\end{eqnarray}

Doing the variable change, $\tau\rightarrow \beta\tau$, in the many-particle partition function, we find
\begin{equation}\label{PF}
Z=\int[d\psi][d\overline{\psi}]e^{ \beta \int_0^1 d\tau \int d^3x {\cal{L}'_\psi}}
\end{equation}
where
\begin{equation}\label{Variable-change}
{\cal{L}'_\psi}=\overline{\psi}\left[\frac{i}{\beta}\gamma^0\partial^0+\mu\gamma^0-i\gamma^1\partial^1-i\gamma^2(\partial^2+ieHx^1)-i\gamma^3\partial^3-m\right]\psi
\end{equation}
Then
\begin{equation}\label{Derivative-PF}
\beta\frac{dZ}{d\beta}=\int[d\psi][d\overline{\psi}]\left\{\beta\int_0^1 d\tau \int d^3x {\cal{L}'_\psi}- \beta\int_0^1 d\tau \int d^3x \overline{\psi}\frac{i\gamma^0\partial^0}{\beta}\psi\right\} e^{\beta  \int_0^1 d\tau \int d^3x {\cal{L}'_\psi}}
\end{equation}
Reversing the variable change (i.e. making $\beta \tau\rightarrow \tau$) we obtain
\begin{equation}\label{Energia}
\langle \widetilde{\tau}_\psi^{00}\rangle=-\beta\left [\frac{(dZ/d\beta)}{Z}-\mu N \right ]=-\frac{\partial\Phi}{\partial T}+\beta\Phi-\beta \mu \frac{\partial\Phi}{\partial\mu}
\end{equation}
where we introduced the grand canonical potential $\Phi=-\frac{1}{\beta}ln Z$, and took into account that $\langle N \rangle=-(\partial\Phi/\partial \mu)_{T,V}$, with $N=\int d^3x \overline{\psi}\gamma^0 \psi$ being the particle-number operator.

Taking into account that the grand canonical potential $\Phi$ is related to the thermodynamic potential $\Omega$ by $\Phi=V\Omega$, and that the system entropy is defined as $S=-\left(\frac{\partial\Omega}{\partial T}\right)_{V,\mu}$, and the particle density by ${\cal{N}}=-\left(\frac{\partial\Omega}{\partial \mu}\right)_{V,T}$, we can add the pure Maxwell energy density to Eq. (\ref{Energia}) to get the system energy density $\varepsilon$ given by
\begin{equation}\label{Energy-Density}
\varepsilon=\frac{1}{\beta V}\langle \widetilde{\tau}_\psi^{00}+ \widetilde{\tau}_M^{00}  \rangle=\Omega_f+TS+\mu {\cal{N}} +\frac{H^2}{2}
\end{equation}

\subsection{Longitudinal Pressure}
As in the energy case, to calculate the pressures we will make use of the functional integral. For the parallel pressure we start by calculating the quantum-statistical average
\begin{equation}\label{T33-average}
\langle \widetilde{\tau}^{33}\rangle=\langle \widetilde{\tau}^{33}_\psi+ \widetilde{\tau}^{33}_{M} \rangle=\frac{\int[d\psi][d\overline{\psi}]\widetilde{\tau}^{33}e^{  \int_0^\beta d\tau \int d^3x {\cal{L}_\psi}(\tau,x)}}{Z}
\end{equation}
where
\begin{equation}\label{T33-average-1}
\widetilde{\tau}^{33}=\int_0^\beta d\tau \int d^3x \tau^{33}(\tau,x)
\end{equation}
and ${\cal{L}_\psi}$ is given in (\ref{Lagrangian}).

For the matter field we have specifically
\begin{equation}\label{T33-average-2}
\langle \widetilde{\tau}^{33}_\psi\rangle=\frac{\int[d\psi][d\overline{\psi}]\widetilde{\tau}_\psi^{33}e^{  \int_0^\beta d\tau \int d^3x {\cal{L}_\psi}}}{Z}
\end{equation}
with $\widetilde{\tau}^{33}_\psi$ given from (\ref{T-Longitudinal}) as
\begin{eqnarray}\label{T33-bar}
\widetilde{\tau}_\psi^{33}=\int_0^\beta d\tau \int d^3x \left(i\overline{\psi}\gamma^3\partial^3\psi\right)+\qquad\qquad\qquad\qquad\qquad\qquad\qquad\quad\nonumber
 \\
 +\int_0^\beta d\tau \int d^3x \overline{\psi}\left[i\gamma^0\partial^0+\mu\gamma^0-i\gamma^1\partial^1-i\gamma^2(\partial^2+ieHx^1)-i\gamma^3\partial^3-m\right]\psi
\end{eqnarray}

Now, making the variable change $x_3\rightarrow Lx_3$ in the partition function, we have
\begin{equation}\label{PF}
Z=\int[d\psi][d\overline{\psi}]e^{  \int_0^\beta d\tau \int d^3x' {\cal{L}'_\psi}}
\end{equation}
where
\begin{equation}\label{Variable-change}
{\cal{L}'_\psi}=\overline{\psi}\left[i\gamma^0\partial^0+\mu\gamma^0-i\gamma^1\partial^1-i\gamma^2(\partial^2+ieHx^1)-\frac{i}{L}\gamma^3\partial^3-m\right]\psi
\end{equation}
with $L$ a scale factor in the $x_3$ direction and $\int d^3x' \equiv L\int d^2x\int_{-1}^1dx_3 = L\int d^3x$.

Then
\begin{equation}\label{Derivative-PF}
L\frac{dZ}{dL}=\int[d\psi][d\overline{\psi}]\left\{L\int_0^\beta d\tau \int d^3x {\cal{L}'_\psi}+ L\int_0^\beta d\tau \int d^3x \overline{\psi}\frac{i\gamma^3\partial^3}{L}\psi\right\} e^{  \int_0^\beta d\tau \int d^3x' {\cal{L}'_\psi}}
\end{equation}
Reversing the variable change (i.e. making $Lx_3\rightarrow x_3$) we obtain
\begin{equation}\label{Average-T33}
\langle \widetilde{\tau}_\psi^{33}\rangle=L\frac{(dZ/dL)}{Z}=-\frac{L}{T}\left(\frac{d\Phi}{dL}\right)
\end{equation}
which can be expressed in terms of the thermodynamic potential as
\begin{equation}\label{Average-T33-2}
\langle \widetilde{\tau}_\psi^{33}\rangle=-\frac{V}{T}\Omega_f
\end{equation}
where $V=LA_\bot$ was considered.

Following a similar procedure for the pure magnetic contribution and taking into account (\ref{Tau}), we find, adding the matter and field contributions,
\begin{equation}\label{Average-T}
\langle \widetilde{\tau}_\psi^{33}+\widetilde{\tau}^{33}_{M}\rangle=-\frac{V}{T}\Omega_f-\frac{V}{T}\frac{H^2}{2}
\end{equation}

Hence, the parallel pressure is given by
\begin{equation}\label{PP-2}
p_\parallel=\frac{1}{\beta V}\langle \widetilde{\tau}_\psi^{33}+\widetilde{\tau}^{33}_{M}\rangle=-\Omega_f-\frac{H^2}{2}
\end{equation}

\subsection{Transverse Pressure}
To find the fermion contribution to the transverse pressure we start from
\begin{equation}\label{Tii-average}
\langle \widetilde{\tau}_\psi^{ii}\rangle=\frac{\int[d\psi][d\overline{\psi}]\widetilde{\tau}_\psi^{ii}e^{  \int_0^\beta d\tau \int d^3x {\cal{L}_\psi}(\tau,x)}}{Z}
\end{equation}
where
\begin{equation}\label{Tii-average-1}
\widetilde{\tau}_\psi^{ii}=\int_0^\beta d\tau \int d^3x \tau_\psi^{ii}(\tau,x),\qquad i=1,2
\end{equation}

The explicit form of the transverse diagonal components of the energy-momentum tensor (\ref{T-Transverse}) in the Landau gauge are given by
\begin{equation}\label{T11-average}
\widetilde{\tau}^{11}_\psi=\overline{\psi}i\gamma^1\partial^1\psi+{\cal{L}}_\psi,\qquad\qquad\quad
\end{equation}
\begin{equation}\label{T22-average}
\widetilde{\tau}^{22}_\psi=\overline{\psi}(i\gamma^2\partial^2-eH\gamma^2x^1)\psi+{\cal{L}}_\psi.
\end{equation}

Apparently, $\tau^{11}$ and $\tau^{22}$ are different, but this is a consequence of the asymmetric Landau gauge we are using. Because the magnetic field is along the $x_3$ axis, there is an $O(2)$ symmetry in the $x_1-x_2$ plane. Hence, the macroscopic pressures, which are obtained after the quantum-statistical average is taken, have to be the same along the $x_1$ and $x_2$ directions ($\langle\widetilde{\tau}^{11}_\psi\rangle=\langle\widetilde{\tau}^{22}_\psi\rangle$). Thus, we can define the transverse pressure as
\begin{equation}\label{P-average-equality}
\langle\widetilde{\tau}^{\bot\bot}_\psi\rangle=\frac{1}{2}(\langle\widetilde{\tau}^{11}_\psi\rangle+\langle\widetilde{\tau}^{22}_\psi\rangle)
\end{equation}
with
\begin{equation}\label{P-average-equality}
\widetilde{\tau}^{\bot\bot}_\psi=\frac{1}{2}\int_0^\beta d\tau \int d^3x \overline{\psi}[i\gamma^1\partial^1+i\gamma^2\partial^2-eH\gamma^2x^1]\psi+S_\psi
\end{equation}
where
\begin{equation}\label{Action}
S_\psi=\int_0^\beta d\tau \int d^3x{\cal{L}}_\psi
\end{equation}

Taking into account that in a uniform magnetic field the transverse motion of charged fermions is quantized in Landau orbits with radii given in units of the magnetic length ${\cal}{l}_H=1/\sqrt{eH}$, we make the variable change
\begin{equation}\label{Variable-change}
x^\bot_i\rightarrow {\cal}{l}_Hx^\bot_i, \qquad x^\bot=(x_1,x_2)
\end{equation}
in the partition function
\begin{equation}\label{PF}
Z=\int[d\overline{\psi}][d\psi]e^{S'_\psi}
\end{equation}
where
\begin{equation}\label{Action'}
S'_\psi={\cal}{l}_H^2\int_0^\beta d\tau \int d^3x\overline{\psi}[i\gamma^\|\cdot\partial_\|-i{\cal}{l}_H^{-1}\gamma^1\partial^1-i{\cal}{l}_H^{-1}\gamma^2\partial^2-{\cal}{l}_H^{-1}\gamma^2x_1-m]\psi
\end{equation}

Taking the derivative of the partition function $Z'$ with respect to the magnetic length, and then reversing the variable change (${\cal}{l}_Hx^\bot_i\rightarrow x^\bot_i$), we obtain
\begin{equation}\label{PF-Derivative}
\frac{{\cal}{l}_H}{2}\frac{dZ}{d{\cal}{l}_H}=\int[d\overline{\psi}][d\psi]\widetilde{\tau}^{\bot\bot}_\psi e^{S_\psi}=\langle \widetilde{\tau}^{\bot\bot}_\psi\rangle \cdot Z
\end{equation}

From where we have
\begin{equation}\label{Tau-Transverse-Average}
\langle \widetilde{\tau}^{\bot\bot}_\psi\rangle = -\frac{\beta {\cal}{l}_H}{2}\frac{d}{d{\cal}{l}_H}(V\Omega_f)
\end{equation}

Taking into account that
\begin{equation}\label{Total-Derv}
\frac{d}{d{\cal}{l}_H}=\frac{\partial}{\partial{\cal}{l}_H}+\left(\frac{\partial H}{\partial {\cal}{l}_H}\right)\frac{\partial}{\partial H}=\frac{\partial}{\partial{\cal}{l}_H}-2H{\cal}{l}_H^{-1}\frac{\partial}{\partial H}
\end{equation}
and that $V=LA_\bot=L\pi {\cal}{l}_H^2$, we can rewrite (\ref{Tau-Transverse-Average}) as
\begin{eqnarray}\label{Tau-Transverse-Average-2}
\langle \widetilde{\tau}^{\bot\bot}_\psi\rangle = -\frac{\beta {\cal}{l}_H}{2}[2L\pi {\cal}{l}_H\Omega_f-V 2H{\cal}{l}_H^{-1}  \frac{\partial\Omega_f}{\partial{\cal}{l}_H}\Omega_f\nonumber
\\
 =-\beta V\Omega_f+\beta V\left( H\frac{\partial\Omega_f}{\partial H}\right)\qquad\qquad
\end{eqnarray}

Similarly to the longitudinal pressure case, the pure magnetic contribution is
\begin{equation}\label{Tii-average-H}
\langle \widetilde{\tau}^{\bot\bot}_M\rangle=V\beta\frac{H^2}{2},\qquad i=1,2
\end{equation}

Finally, we can find the transverse pressure adding the matter contribution (\ref{Tau-Transverse-Average-2}) and the field contribution (\ref{Tii-average-H}), as
\begin{equation}\label{TP}
p_\bot=\frac{1}{\beta V} \langle \widetilde{\tau}_\psi^{\bot\bot}+\widetilde{\tau}_M^{\bot\bot}\rangle=-\Omega_f-HM_f+\frac{H^2}{2}
\end{equation}
where $M_f=-(\partial\Omega_f/\partial H)$ is the fermion-system magnetization.

\subsection{Energy-Momentum Tensor Covariant Structure for the Magnetized Fermion System}

In analogy to Eq. (\ref{Tau}) for the energy-momentum tensor of the magnetic field, we give here a covariant decomposition for the energy momentum tensor of the whole system containing the matter and field contributions. In this form, we summarize the results for the energy density and pressures given in Eqs. (\ref{Energy-Density}), (\ref{PP-2}) and (\ref{TP}). In order to accomplish this goal, we define the system thermodynamic potential as the sum of the matter and field contributions
\begin{equation}\label{System-Omega}
\Omega=\Omega_f+\frac{H^2}{2}
\end{equation}

Then, we have
\begin{equation}\label{System-T}
\frac{1}{\beta V}\langle \widetilde{\tau}^{\mu\nu}\rangle=\Omega \eta^{\mu\nu}+(\mu {\cal{N}}+TS)u^{\mu}u^{\nu}+HM\eta^{\mu\nu}_\perp,
\end{equation}
where $\eta^{\mu\nu}_\perp$ was defined in (\ref{Tau}), and $M=-(\partial\Omega/\partial H)$.

In the quantum field limit, i.e. when $T=\mu=H=0$, the only term different from zero is the first one in the RHS of (\ref{System-T}). In that case the system has Lorentz symmetry. If temperature and/or density are switched on, then the Lorentz symmetry is broken specializing a particular reference frame comoving with the medium center of mass and having a four velocity $u_\mu=(1,\overrightarrow{0})$. This is reflected in the second term of the RHS of (\ref{System-T}). Finally, when there is an external uniform magnetic field acting on the system, the additional symmetry breaking $O(3)\rightarrow O(2)$ takes place, and $\langle \widetilde{\tau}^{\mu\nu}\rangle$ get an anisotropy reflected in the appearance of the transverse metric structure $\eta^{\mu\nu}_\perp$ in (\ref{System-T}).

\section{Equation of State}\label{EoS}
As seen from Eqs. (\ref{Energy-Density}), (\ref{PP-2}) and (\ref{TP}), the system energy density and pressures depend on the fermion system thermodynamic potential $\Omega_f$. For the dense ($\mu \neq 0$) fermion system under an applied uniform magnetic field with Lagrangian density (\ref{Lagrangian}) the thermodynamic potential is given by (See Appendix \ref{Thermo-Pot})
\begin{equation}\label{Therm-Potential}
\Omega_f(\mu, T, H)=-\frac{qH}{2\pi^2}\sum_{n=0}^\infty d(n)\int_0^\infty dp_3  \left[\varepsilon_n  +\frac{1}{\beta}ln \left( 1+e^{-\beta(\varepsilon_n-\mu)}\right)\left(1+e^{-\beta(\varepsilon_n+\mu)}\right)  \right]
\end{equation}
where $d(n)=2-\delta_{n0}$ is the spin degeneracy of the $n$ Landau level, and $\varepsilon_n=\sqrt{p_3^2+m^2+2qHn}$ is the energy of the particle in the $n$ Landau level with $n=0,1,2,...$. In the bracket of the RHS of Eq. (\ref{Therm-Potential}) the first term is the QFT contribution which is independent on the temperature $T$ and chemical potential $\mu$, and the second term is the statistical contribution depending on these two parameters.

The particle number density can be obtained from (\ref{Therm-Potential}) as
\begin{equation}\label{Particle-Density-1}
{\cal{N}}=-\frac{\partial\Omega_f}{\partial \mu} = \frac{qH}{2\pi^2}\sum_{n=0}^\infty d(n) \int_0^\infty dp_3\left[\frac{1}{1+e^{\beta(\varepsilon_n-\mu)}}-\frac{1}{1+e^{\beta(\varepsilon_n+\mu)}}\right]
\end{equation}

We are interested in systems, as neutron stars, where the leading parameter is the fermion density $\mu$. Neutron stars cool rapidly through neutrino emission. They can reach in a few seconds temperatures $T\lesssim 10^8 K$ \cite{Temperature-1}. Thus, in applications to astrophysical compact objects, it is considered that the stellar medium is highly degenerate, so usually neglecting the thermal effects. Then, to find the zero temperature limit ($\beta\rightarrow \infty$) of the thermodynamic potential, we will consider only the statistical part in (\ref{Therm-Potential}),
\begin{equation}\label{Therm-Potential-0}
\Omega_f^0=\frac{qH}{2\pi^2}\sum_{n=0}^{n_H} d(n)\left[\int_0^{\sqrt{\mu^2 - a_n^2}} dp_3 \sqrt{p_3^2+a_n^2}-\mu \sqrt{\mu^2-a_n^2} \right],
\end{equation}
where $a_n^2=m^2+2qHn$ and $n_H=I[(\mu^2-m^2)/2qH]$, with $I[...]$ denoting the integer part of the argument. The zero temperature limit of the particle number density (\ref{Particle-Density-1}) is
\begin{equation}\label{Particle-Density}
{\cal{N}}^0= \frac{qH}{2\pi^2}\sum_{n=0}^{n_H} d(n) \sqrt{\mu^2-a_n^2}
\end{equation}

The system magnetization depending on the chemical potential can be found from Eq. (\ref{Thermo-Potential-4}) in the Appendix as
\begin{eqnarray}\label{Magnetization-T}
M=-\frac{\partial\Omega_f^{SQFT}}{\partial H}=\frac{q}{2\pi^2\beta}\sum_{n=0}^\infty d(n)\int_0^\infty dp_3 ln \left[\left( 1+e^{-\beta(\varepsilon_n-\mu)}\right)\left(1+e^{-\beta(\varepsilon_n+\mu)}\right)\right]\nonumber
\\
-\frac{q}{2\pi^2}\sum_{n=0}^\infty d(n)\int_0^\infty dp_3 \frac{qHn}{\varepsilon_n}\left[\frac{1}{1+e^{\beta(\varepsilon_n-\mu)}}+\frac{1}{1+e^{\beta(\varepsilon_n+\mu)}}\right]
\end{eqnarray}
and in the zero-temperature limit it takes the form
\begin{equation}\label{Magnetization-T0}
M^0= -\frac{q}{2\pi^2}\sum_{n=0}^{n_H} d(n)\int_0^{\sqrt{\mu^2-a^2}} dp_3 (\varepsilon_n-\mu+\frac{qHn}{\varepsilon_n})
\end{equation}

The EoS of the dense magnetized system at zero temperature will be given by the inter-relation of the energy density, and the parallel and transverse pressures at zero temperature, which can be obtained from (\ref{Energy-Density}), (\ref{PP-2}), (\ref{TP}), (\ref{Therm-Potential-0}), (\ref{Particle-Density}), and (\ref{Magnetization-T0}), as
\begin{equation}\label{EoS-T0}
\varepsilon^0=\Omega^0+\mu {\cal{N}}^0, \qquad  {p_\parallel}^0=-\Omega^0, \qquad {p_\bot}^0=-\Omega^0+H\frac{\partial\Omega^0}{\partial H}
\end{equation}
where $\Omega^0$ is the general thermodynamic potential of the system, given by $\Omega^0=\Omega_f^0+ \frac{H^2}{2}$.

In the case of quark matter, the asymptotically-free phase of quarks will form a perturbative vacuum (inside a bag) which is immersed in the nonperturbative vacuum. This scenario is what is called the MIT bag model \cite{Bag-Model}. The creation of the "bag" costs free energy. Then, in the energy density $\Omega^0$, the energy difference between the perturbative vacuum and the true one should be added. Essentially, that is the bag constant $B$ characterizing a constant energy per unit volume associated to the region where the quarks live. From the point of view of the pressure, $B$ can be interpreted as an inward pressure needed to confine the quarks into the bag. Hence, in the case of quark matter, Eqs. (\ref{EoS-T0}) get extra terms depending on $B$,
\begin{equation}\label{EoS-T0-2}
\varepsilon^0=\Omega_f^0+\mu {\cal{N}}^0+ \frac{H^2}{2}+B, \qquad  {p_\parallel}^0=-\Omega_f^0-\frac{H^2}{2}-B, \qquad {p_\bot}^0=-\Omega_f^0-HM^0+\frac{H^2}{2}-B
\end{equation}
Because the relative sign between $B$ and the magnetic energy $H^2/2$ term is not the same for the parallel and transverse pressures, the pure magnetic energy contribution cannot be absorbed by the vacuum energy $B$. Taking into account that the magnetic field varies in several orders from the inner core to the star surface, the term $H^2/2$ applies a tremendous extra pressure on the quark matter, but because of the anisotropy between the longitudinal and transverse directions with respect to the field alignment, the pressure coming from $\frac{H^2}{2}$ is negative on the parallel pressure, while on the transverse pressure it is positive.

\begin{figure}
\begin{center}
\includegraphics{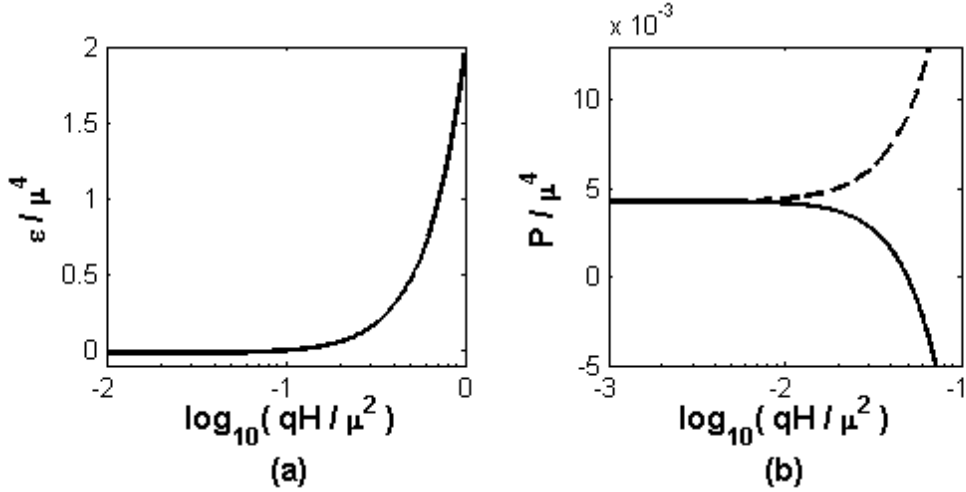}
\caption{(a) System energy density vs magnetic field for $\mu=400 MeV$. (b) System pressures (parallel pressure represented in solid line, and transverse pressure in dash line) vs magnetic field for $\mu=400 MeV$.} \label{fig2}
\end{center}
\end{figure}

The phenomenological parameter $B$ is estimated taking into account the underlaying dynamics and external conditions (as temperature and density) of the system \cite{Variable-B}, but it cannot be calculated from first principles due to our present limitations in dealing with non-perturbative QCD. In our analysis of the magnetic field dependence of the equation of state, we will only consider the free-fermion case, hence we take $B=0$.

In Fig. 2, we are showing how the energy density and pressures change with the magnetic field at fixed $\mu$. We found that the energy density and transverse pressure increase with the magnetic field, while the parallel pressure decreases. Also we have that, in our free-fermion model, for field strengths close to $10^{19} G$ the parallel pressure becomes negative. Hence, field strengths of that order can produce strong instabilities in the star's structure. It will be interesting to investigate this issue in more realistic models that include particle interactions.

\begin{figure}
\begin{center}
\includegraphics{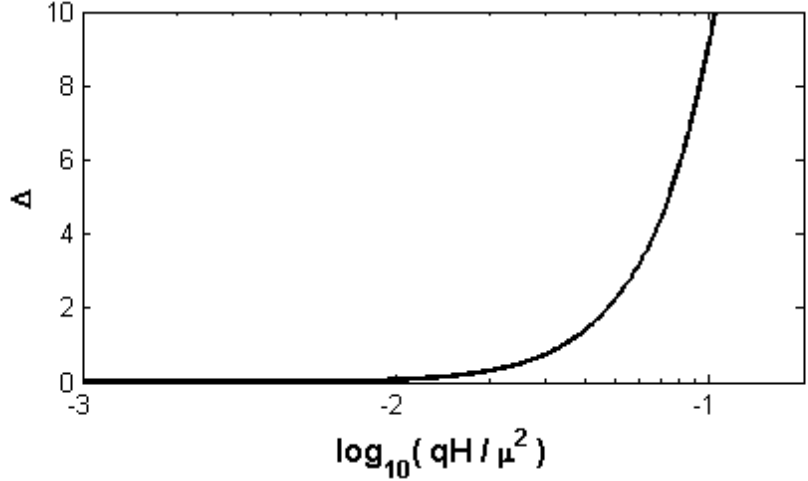}
\caption{Splitting coefficient $\Delta$ vs magnetic field for $\mu=400 MeV$} \label{fig3}
\end{center}
\end{figure}
The splitting between the two pressures relative to their weak field value\footnote{Notice in Fig. 2 that at weak field ($eH\ll\mu^2$) the two pressures coincide.}
\begin{equation}\label{Pressure-Splitting-ratio}
\Delta=\frac{\mid{p_\bot}^0-{p_\parallel}^0\mid}{\mid{p_\parallel}^0(eB\ll \mu^2)\mid}
\end{equation}
is given in Fig. 3. There, we can see that for $\mu=400MeV$, fields $H\geq 10^{18}G$ have splitting rates $\Delta\geq 10$. This result indicates that for the field-strength range that can take place in the inner core of magnetars, for example, the pressure anisotropy can play an important role in the star structure and geometry. A criterium to separate the isotropic regime (${p_\bot}_0={p_\parallel}_0$) from the anisotropic one can be
\begin{equation}\label{Criterium}
\Delta\simeq {\cal{O}}(1)
\end{equation}
For the density value previously considered ($\mu=400 MeV$), we have a threshold field of order $10^{17} G$.

\begin{figure*}
\includegraphics{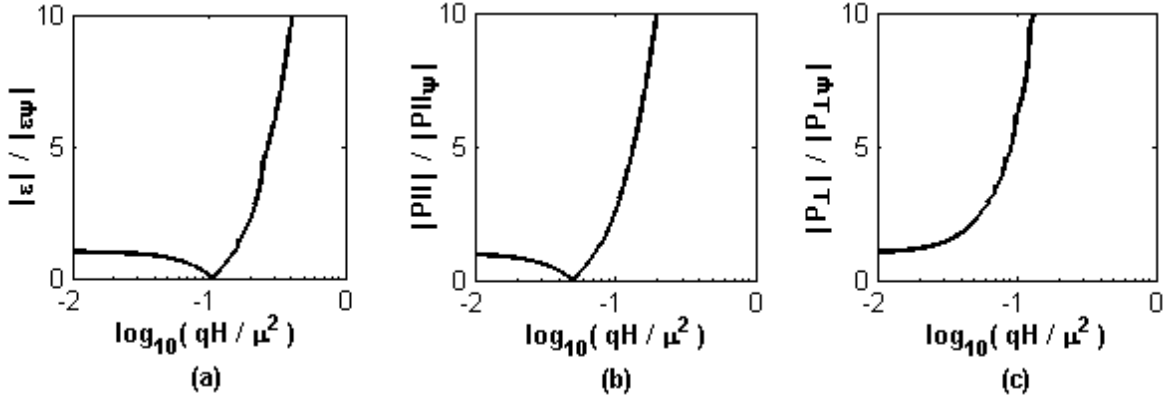}
\caption{(a) Ratios of system/matter energy density, (b) parallel pressure, and (c) transverse pressure vs magnetic field for $\mu=400 MeV$.} \label{fig4}
\end{figure*}

To determine for what field values the pure magnetic contribution to the energy density and pressures becomes important, we plotted the ratio between the total energy (i.e. the one having the matter and field contributions) and the matter energy density in Fig. 4-(a); that of the total parallel pressure to the parallel matter pressure in Fig. 4-(b); and the corresponding one to the transverse pressure in Fig. 4-(c). From those graphs we have that the pure magnetic contribution becomes significant for field strengths between one and two orders smaller than $\mu^2$. Thus, for $\mu=400 MeV$, fields larger than $10^{17} G$ will make a significant contribution in the system energy and pressures through the $H^2/2$ term. Hence, in astrophysical applications where the field strength in the inner core of compact objects can be large enough, the pure field contribution should be consider on equal footing as the matter contribution in determining the parameters of the EoS. It is worth noticing that the pure magnetic contribution becomes significant for fields of the order of the threshold field for the isotropic-anisotropic transition. This is not a coincidence, but a consequence of the fact that the main contribution to the pressure splitting
\begin{equation}\label{Pressure-Splitting}
{p_\bot}^0-{p_\parallel}^0=HM^0+H^2
\end{equation}
comes from their magnetic term $H^2$. The term depending on the magnetization $M^0$ has an oscillatory behavior due to the
Haas-van Alphen oscillations appearing at low temperature in degenerate fermionic systems \cite{HvA} (See Fig. 5). Nevertheless, the Haas-van Alphen oscillations are not noticeable in Fig. 3 because their amplitudes are much smaller than $H$, so they can be neglected in (\ref{Pressure-Splitting}).

\begin{figure}
\includegraphics{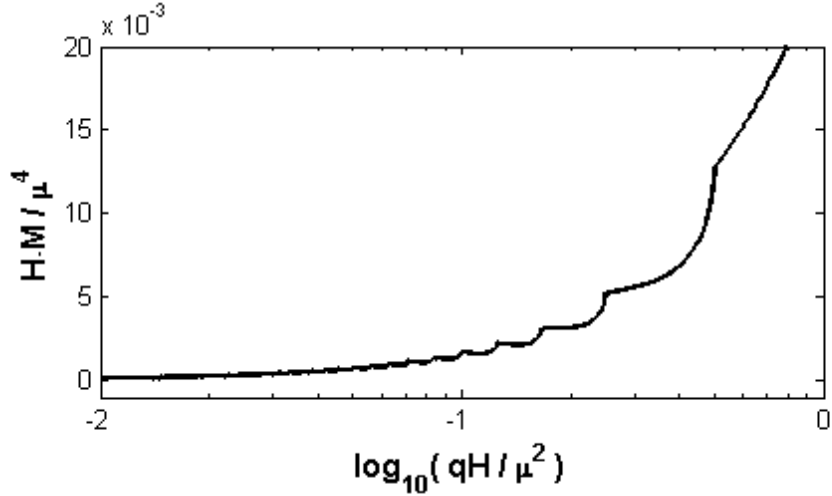}
\caption{System magnetization $M$ times the magnetic field $H$ vs the magnetic field for $\mu=400 MeV$.} \label{fig5}
\end{figure}

The threshold value of $\sim 10^{17} G$ we found, is model dependent. It is applicable to systems of free fermions under certain parameter values. For other system, such as cold dense quark systems with superconducting gaps, the corresponding threshold field should be determined.

Finally, in Fig. 6 the system EoS is plotted . There, the variation of the energy density vs the parallel pressure is given in Fig. 6-(a), and with respect to the transverse pressure in Fig. 6-(b), for a fixed density ($\mu=400 MeV$) and a variable magnetic field. Due to the pressure anisotropy, the EoS in this case should be given by a curve in a three dimensional representation with axes $(\varepsilon,p_\|,p_\bot)$. In Fig. 6 we give the projections of that curve in the two planes, both including the $\varepsilon$ axis.

\begin{figure}
\includegraphics{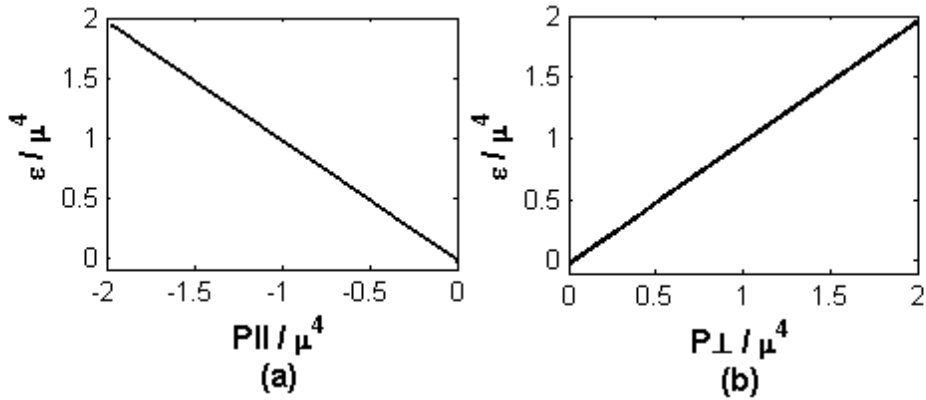}
\caption{(a) System energy density minus the bag constant $B$ vs the parallel pressure and (b) vs the transverse pressure, for magnetic fields in the range $0<qH/\mu^2<1$ and $\mu=400 MeV$.} \label{fig6}
\end{figure}

\section{Concluding Remarks}\label{conclusions}
In this paper we have shown that a relativistic dense fermion gas under a sufficiently strong magnetic field has a highly anisotropic EoS. For field strengths about two orders smaller than $\mu^2$, the anisotropy effects begin to emerge, and once the field reaches values one order smaller than  $\mu^2$, they cannot be neglected anymore, since the splitting is ten times the value of the pressure at zero field. The splitting of the pressure in two terms, one along the field direction (the parallel pressure) and the other, perpendicular to the field direction (the transverse pressure), should be taken into account in astrophysical considerations when studying compact objects that exhibit strong magnetic fields, as it may affect the structure and geometry of the star.

At strong magnetic fields ($H\sim \mu^2/10$), the pure magnetic energy contribution $\varepsilon_M$, as well as the magnetic pressures ${p_\|}_M$ and ${p_\bot}_M$, were found to be as important as the corresponding matter contributions $\varepsilon_f$, ${p_\|}_f$ and ${p_\bot}_f$ (see Fig. 5 for details). Therefore, a sufficiently high magnetic field may actually influence the EoS in two different ways: by modifying the matter contributions to the energy density and pressures, and as importantly, through the pure magnetic (coming from the Maxwell term in the original Lagrangian) contribution to the energy and pressures. The order of the fields required for the two contributions to become comparable was found to be the same as the field strength needed for the pressure anisotropy to be relevant. We call attention that the magnetic contribution into the EoS has been proved to have significant effects in the magnetohydrodynamics of quark-gluon plasma \cite{Neda}.

Our results are valid for relativistic systems of fermions in the presence of a uniform and constant magnetic field. These kinds of systems have been considered in the physics of neutron stars, as well as quark stars. At present, it is known that if the density is high enough to produce a quark deconfined phase in the star core, the state that minimizes the energy of such system will be a color superconductor \cite{Bailin-Love}. There exist already several papers where the effect of the color superconducting gap was considered in the energy density and pressure of highly dense quark matter \cite{EP-Gap}.

In spin-zero color superconductivity, the color condensates in the CFL, as well as in the 2SC phases,
have non-zero electric charge. Then, we could expect that the photon acquires a Meissner mass which produces the screening of a weak magnetic field (the well known phenomenon of Meissner effect). Nevertheless, in the spin-zero color superconductor the conventional electromagnetic field $A_\mu$ is not an eigenfield, but it is mixed with the $8^{th}$-gluon $G^8_\mu$ to form a long-range field that becomes the in-medium electromagnetic field $\widetilde{A}_\mu$ (i.e. the so called "rotated" electromagnetic field) \cite{alf-raj-wil-99/537}. In a series of papers \cite{MCFL} there has been shown that the color-superconducting properties of a three-flavor
massless quark system are substantially
affected by the penetrating "rotated" magnetic field and as a
consequence, a new phase, called Magnetic Color Flavor Locked (MCFL)
phase, takes place. In the MCFL phase the gaps that receive contributions from pairs of charged quarks get reinforced at very strong fields producing a sizable splitting as compared with the gaps that only get contribution from pairs of neutral quarks. As the field decreases, the gaps become oscillating functions of the magnetic field \cite{Oscillations}, a phenomenon associated with the known Haas-van Alphen oscillations appearing in magnetized systems \cite{HvA}.

Therefore, a realistic study of the EoS of stellar magnetized dense quark matter should be formalized in the color superconducting MCFL phase. A preliminary study \cite{Aurorita} already exists, where the magnetic field effect was considered only in the quark particle spectrum, but no in the gap, which was taken as a fixed parameter. In a future publication \cite{EoS-CS}, a self-consistent investigation of the EoS in the MCFL phase, taking into account the gap equations depending on the magnetic field, will be presented.

{\bf Acknowledgments:} This work was supported in part by the Office of Nuclear Physics of the Department of Energy under contract DE-FG02-09ER41599.

\appendix

\section{Thermodynamic Potential for a Dense-Magnetized Fermion System}
\label{Thermo-Pot}
The thermodynamic potential, $\Omega_f$, in the presence of a constant and uniform magnetic field has been previously calculated using different methods at finite temperature and/or chemical potential. For example, using the Schwinger proper time method \cite{Proper-Time}, $\Omega_f$ was calculated in Refs. \cite{Temperature} and \cite{Chem-Pot}, at $T\neq 0$, and introducing a chemical potential $\mu\neq 0$, respectively. In the Furry picture \cite{Furry}, $\Omega_f$ was calculated at $T\neq 0$ and $\mu\neq 0$ in \cite{Real-Time}, and using the worldline method at $T\neq 0$ in \cite{World-Line}.

Here we present in detail the calculation of the thermodynamic potential in the presence of a constant and uniform magnetic field at $T\neq 0$ and $\mu \neq 0$ using Ritus's method. This approach was originally developed for charged spin-1/2 particles \cite{Ritus:1978cj}, and later on extended to charged spin-1 bosons \cite{efi-ext}. Recently, it has been implemented for the case of spin-1/2 in an inhomogeneous magnetic field in reduced dimensions \cite{Raya}. This approach is
based on a Fourier-like transformation performed by the eigenfunction matrices $E_p(x)$ which are associated to the wave functions of the asymptotic states of charged
fermions in a uniform magnetic field. The $E_p(x)$ functions play the role in the
magnetized medium of the usual plane-wave (Fourier) functions $e^{i
px}$ at zero field. The advantage of this method is that the field-dependent fermion Green function is diagonalized in momentum space, so having a similar form to that in the free space. Hence, this formalism is very convenient to implement the statistical sum by the imaginary time procedure needed to describe systems at finite temperature and density . Also, the obtained Green function in momentum space explicitly depends on the Landau levels. This last result makes it particularly convenient to be used in the strong-field approximation, where one can
constraint the calculations to the contribution of the lowest Landau level
(LLL) in the particle spectrum.

 Ritus's method has been successfully used in the context of
chiral symmetry breaking in a magnetic field \cite{orthonormality}, as well as in magnetized color superconductivity \cite{MCFL}, \cite{M2SC}.

The thermodynamic potential of the magnetized dense system at finite temperature $\Omega_f(H,\mu,T)$ is given by
\begin{equation}\label{Therm-Potential-Def}
\Omega_f(H,\mu,T)=\frac{\Phi(H,\mu,T)}{V},
\end{equation}
where $\Phi(H,\mu,T)$ is the grand canonical potential (in functional terminology, the effective action in the presence of an external magnetic field at finite temperature and density), which is given in terms of the inverse fermion propagator as
\begin{equation}\label{Grand-Potential-Def}
\Phi(H,\mu,T)=\frac{1}{\beta} Tr \ln Z=\frac{i}{\beta} Tr \ln G^{-1}(x,x'),
\end{equation}
with the trace and logarithm taken in a functional sense, and $G^{-1}(x,x')$ being the fermion inverse propagator in space representation. To make the transformation to momentum space, because of the dependence of $G^{-1}(x,x')$ on the electromagnetic potential of the external field $A_{\mu }^{\mathit{ext}} = H x_1 \delta_{\mu 2}$, it is convenient to use the Ritus's transformation
\begin{eqnarray}\label{Propagator}
G^{-1}(x,x')= \sum\hspace{-0.49cm}\int \frac{d^{4}p^E}{\left( 2\pi
\right) ^{4}}
E_{p}^{l}(x)\Pi(l)\widetilde{G}^{-1}_l(\overline{p})\overline{E}_{p}^{l}(x')
\end{eqnarray}
where $\sum_{\it l}\hspace{-0.47cm}\int \frac{d^{4}p^E}{\left( 2\pi
\right) ^{4}}=i\sum_{l=0}^\infty\frac{dp_4dp_2dp_3}{(2\pi)^4}$, and we introduced the Ritus' transformation functions
\begin{equation}\label{Ep}
 E_{p}^{l}(x)=E_{p}^{+}(x)\Delta(+)+E_{p}^{-}(x)\Delta(-)
\end{equation}
with
\begin{equation}
\Delta(\pm)=\frac{I\pm i\gamma^{1}\gamma^{2}}{2},
\label{Spin-projectors}
\end{equation}
representing the spin up ($+$) and down ($-$) projectors, and
$E_{p}^{+/-}(x)$ are the corresponding eigenfunctions
\begin{eqnarray}\label{E-x}
E_{p}^{+}(x)=N_{l}e^{-i(p_{0}x^{0}+p_{2}x^{2}+p_{3}x^{3})}D_{l}(\rho),\qquad
\nonumber
\\
E_{p}^{-}(x)=N_{l-1}e^{-i(p_{0}x^{0}+p_{2}x^{2}+p_{3}x^{3})}D_{l-1}(\rho)
\end{eqnarray}
with normalization constant $N_{l}=(4\pi eH)^{1/4}/\sqrt{l!}$,
$D_{l}(\rho)$ denoting the parabolic cylinder functions of argument
$\rho=\sqrt{2eH}(x_{1}-p_{2}/eH)$, and index given by the Landau
level numbers $l=0,1,2,...$.

The $E_p^l$ functions satisfy the orthogonality condition
\cite{Wang}
\begin{equation}
\int d^{4}x \overline{E}_{p}^{l}(x)E_{p'}^{l'}(x)=(2\pi)^4
\widehat{\delta}^{(4)}(p-p')\Pi(l) \ , \label{orthogonality}
\end{equation}
with $\overline{E}_{p}^{l}\equiv \gamma^{0}
(E_{p}^{l})^{\dag}\gamma^{0}$, $\Pi(l)=\Delta(+)\delta^{l0}+I(1-\delta^{l0})$, and $\widehat{\delta}^{(4)}(p-p')=\delta^{ll'} \delta(p_{0}-p'_{0})
\delta(p_{2}-p'_{2}) \delta(p_{3}-p'_{3})$.

The spin structure of the $E_p$ functions is essential to satisfy
the eigenvalue equations
\begin{equation}
(\Pi\cdot\gamma)E^{l}_{p}(x)=E^{l}_{p}(x)(\gamma\cdot\overline{p}) \
, \label{eigenproblem}
\end{equation}
with $\overline{p}^\mu=(p^{0},0, -\sqrt{2eH l},p^{3})$.

Using (\ref{orthogonality}) and (\ref{eigenproblem}), the inverse propagator in momentum representation $\widetilde{G}^{-1}_l(\overline{p})$ appearing in (\ref{Propagator}), is found from
\begin{eqnarray}\label{Inv-Propagator}
G^{-1}_{l}(p,p')= \int d^4xd^4y
\overline{E}_{p}^{l}(x)[\Pi_\nu \gamma^\nu+\mu \gamma^0-m]E_{p'}^{l'}(y)=\nonumber
 \\
=(2\pi)^4\widehat{\delta}^{(4)}(p-p')\Pi(l)\widetilde{G}^{-1}_l(\overline{p})\quad \quad \quad \quad \quad \quad
\end{eqnarray}
where
\begin{equation}
\Pi_{\mu}=i\partial_{\mu}-eA_{\mu} \label{9}
\end{equation}
and
\begin{equation}
\widetilde{G}^{-1}_l(\overline{p})=[{\overline{p}}^{*}\cdot\gamma
-m] \label{Inv-Propagator-Def}
\end{equation}
with ${\overline{p}}^{*}_\nu=(ip^{4}-\mu,0, \sqrt{2eH l},p^{3})$

Substituting (\ref{Propagator}) in (\ref{Grand-Potential-Def}), taking the functional trace, and using he orthogonality condition (\ref{orthogonality}), we obtain
\begin{eqnarray}\label{Grand-Potential-2}
\Phi(H,\mu,T)=
\frac{i}{T} tr \ln \int dx \int dx'\delta^4(x,x')\sum\hspace{-0.49cm}\int \frac{d^{4}p}{\left( 2\pi
\right) ^{4}}
E_{p}^{l}(x)\Pi(l)\widetilde{G}^{-1}_l(\overline{p})\overline{E}_{p}^{l}(x')=\nonumber
 \\
 = \frac{i\widehat{\delta}^3_p(0)}{T}tr \ln \sum\hspace{-0.49cm}\int \Pi(l)\widetilde{G}^{-1}_l(\overline{p})d^{3}\widehat{p}\qquad \qquad \qquad \qquad \qquad \qquad \qquad \quad
\end{eqnarray}
In (\ref{Grand-Potential-2}), $d^{3}\widehat{p}=dp_0dp_2dp_3$, and $tr$ denotes the remaining spinorial trace. Now, taking into account that
\begin{equation}
\widehat{\delta}^3_p(0)=\frac{1}{(2\pi)^3}\int_0^\beta dx_4\int_{-\infty}^{\infty} dx_2dx_3\label{Delta-zero-Def}
\end{equation}
and because $\widetilde{G}^{-1}_l(\overline{p})$ does not depend on $p_2$, for the integration in $p_2$ in (\ref{Grand-Potential-2}) we have
\begin{equation}
\int_{-\infty}^\infty\frac{dp_2}{2\pi}=\int_{-\infty}^{\infty}\frac{dp_2}{2\pi}  e^{-i\frac{p_2p_1}{eH}} |_{p_1=0} =\frac{1}{l_{H}^2}\widehat{\delta}_{p_1}(0)
=\frac{eH}{2\pi}\int_{-\infty}^{\infty} dx_1  \label{Int-p2}
\end{equation}

Substituting (\ref{Delta-zero-Def}) and (\ref{Int-p2}) in (\ref{Grand-Potential-2}), we obtain in Euclidean space $(p_0\rightarrow ip_4)$,
\begin{equation}\label{Grand-Potential-3}
\Phi(H,\mu,T)=-eHV\beta tr \ln\sum_{l=0}^{\infty}\int_{-\infty}^{\infty}\frac{dp_4dp_3}{(2\pi)^3} \Pi(l)\widetilde{G}^{-1}_l(\overline{p})
\end{equation}

Taking into account that $\Pi(l)$ separates the $l=0$ Landau level from the rest, substituting (\ref{Grand-Potential-3}) in (\ref{Therm-Potential-Def}), and because of the identity $tr \ln \widehat{O}=\ln \det \widehat{O}$, we obtain for the thermodynamic potential
\begin{equation}\label{Grand-Potential-4}
\Omega_f(H,\mu,T)= -eH\left[\int_{-\infty}^{\infty}\frac{dp_4dp_3}{(2\pi)^3} \ln \det \widetilde{G}^{-1}_0(\overline{p})+ 2\sum_{l=1}^{\infty}\int_{-\infty}^{\infty}\frac{dp_4dp_3}{(2\pi)^3} \ln \det \widetilde{G}^{-1}_l(\overline{p})\right]
\end{equation}

The fermion system at finite temperature can be described by taking the discretization of the fourth momentum following Matsubara's procedure
 \begin{equation}\label{Matsubara}
\int_{-\infty}^\infty\frac{dp_4}{2\pi}\rightarrow\frac{1}{\beta}\sum_{p_4}, \qquad p_4=\frac{(2n+1)\pi}{\beta}, \quad n=0,\pm1,\pm2,...
\end{equation}

Taking into account (\ref{Inv-Propagator-Def}), after taking the trace in (\ref{Grand-Potential-4}) we obtain
\begin{equation}\label{Thermo-Potential}
\Omega_f(H,\mu,T)=-\frac{eH}{\beta}\int_{-\infty}^\infty \frac{dp_3}{4\pi^2}\sum_{l=0}^{\infty}(2-\delta_{l0})\sum_{p_4}\ln \left[(p_4+i\mu)^2+\epsilon_l^2\right]
\end{equation}
where
\begin{equation}\label{Disp-Rel}
\epsilon_l^2=p_3^2+2|eH|l+m^2
\end{equation}

Notice that in this approach the sum in the $p_4$ term, which is obtained in (\ref{Thermo-Potential}), is formally similar to that appearing in the free-particle thermodynamic potential (i.e. at $H=0$ and $\mu\neq 0$) \cite{Kapusta}. After summing in $p_4$ we get
\begin{equation}\label{Thermo-Potential-2}
\Omega_f(H,\mu,T)=-\int_{-\infty}^\infty dp_3\sum_{l=0}^{\infty}\frac{eHd(l)}{4\pi^2}\left[\epsilon_l+\frac{1}{\beta}\ln \left(1+e^{-\beta(\epsilon_l-\mu)}\right)\left(1+e^{-\beta(\epsilon_l+\mu)}\right)\right]
\end{equation}
The ratio $\frac{eHd(l)}{4\pi^2}$, with $d(l)=2-\delta_{l0}$, is the density of states per Landau level. The factor $d(l)$ is the spin degeneracy of Landau levels with $l\neq 0$.

The thermodynamic potential (\ref{Thermo-Potential-2}) has two contributions. One that does not depend on the temperature and chemical potential $\Omega^{QFT}$, and the statistical one $\Omega^{SQFT}$, given respectively by
\begin{equation}\label{Thermo-Potential-3}
\Omega_f^{QFT}(H)=-\frac{eH}{4\pi^2}\int_{-\infty}^\infty dp_3\sum_{l=0}^{\infty}d(l)\epsilon_l
\end{equation}
\begin{equation}\label{Thermo-Potential-4}
\Omega_f^{SQFT}(H,T,\mu)=-\frac{eH}{4\pi^2\beta}\int_{-\infty}^\infty dp_3\sum_{l=0}^{\infty}d(l)\ln \left(1+e^{-\beta(\epsilon_l-\mu)}\right)\left(1+e^{-\beta(\epsilon_l+\mu)}\right)
\end{equation}
As known, $\Omega^{QFT}(H)$ has non-field-dependent ultraviolet divergencies that should be renormalized (see Ref. \cite{Lifshitz} for a detailed renormalization procedure of this term). After renormalization the well-known Schwinger expression \cite{Proper-Time}
\begin{equation}\label{Thermo-Potential-5}
\Omega_f^{QFT}(H)=-\frac{1}{8\pi^2}\int_{0}^\infty \frac{ds}{s^3}exp(-m^2s)\left(esH\coth (esH)-1-\frac{1}{3}(esH)^2 \right)
\end{equation}
is found. In the calculations of the energy density and pressures done in Sec. V we considered only the $\Omega^{SQFT}$ contribution, since for astrophysical applications one always has $\mu^2\gg eH$, and the leading contribution of $\Omega_f$ will come from $\Omega_f^{SQFT}$.

\end{document}